\documentclass[aps,prd,preprint,groupedaddress,nofootinbib]{revtex4-1}
\usepackage{amsmath,latexsym,amssymb,amsfonts}
\usepackage{float} 
\usepackage{graphicx}% Include figure files
\usepackage{epstopdf}
\usepackage{graphics}
\usepackage{caption}
\usepackage{subfigure}
\usepackage{bm}
\usepackage{color}
\usepackage{xcolor}
\usepackage{hyperref}
\usepackage[dvips]{psfrag}
\usepackage{rotating}
\usepackage{soul}
\usepackage[normalem]{ulem}

\begin{document}
\title{\boldmath Primordial black holes and induced gravitational waves from inflation in the Horndeski theory of gravity}
\author{ 
Pisin Chen,$^{1, 2, 3,}$\footnote{pisinchen@phys.ntu.edu.tw} 
Seoktae Koh,$^{4,}$\footnote{kundol.koh@jejunu.ac.kr}
and Gansukh Tumurtushaa$^{1,}$\footnote{gansuh@ntu.edu.tw} 
}

\affiliation{$^{1}$\small{Leung Center for Cosmology and Particle Astrophysics, National Taiwan University, Taipei 10617, Taiwan, ROC}\\
$^{2}$\small{Department of Physics and Center for Theoretical Sciences, National Taiwan University, Taipei 10617, Taiwan, ROC}\\
$^{3}$\small{Kavli Institute for Particle Astrophysics and Cosmology, SLAC National Accelerator Laboratory, Stanford University, Stanford, California 94305, USA}\\
$^{4}$\small{Department of Science Education, Jeju National University, Jeju, 63243, Korea}}

%\date{\today}

\begin{abstract}
We investigate the production of primordial black holes (PBHs) and scalar-induced gravitational waves (GWs) for cosmological models in the Horndeski theory of gravity. The cosmological models of our interest incorporate the derivative self-interaction of the scalar field and the kinetic coupling between the scalar field and gravity. We show that the scalar power spectrum of the primordial fluctuations can be enhanced on small scales due to these additional interactions. Thus, the formation of PBHs and the production of induced GWs are feasible for our model. Parameterizing the scalar power spectrum with a local Gaussian peak, we first estimate the abundance of PBHs and the energy spectrum of GWs produced in the radiation-dominated era. Then, to explain the small-scale enhancement in the power spectrum, we reconstruct the inflaton potential and self-coupling functions from the power spectrum and their spectral tilt. Our results show that the small-scale enhancement in the power spectrum can be explained by the local feature, either a peak or dip, in the self-coupling function rather than the local feature in the inflaton potential.
\end{abstract}

%\arxivnumber{xxxx.xxxxx}

\maketitle
%\flushbottom
%\tableofcontents

%---------------------------------------------------------------------------------------------------------------------------------------------------------------------------------------------------------------------------------------
\section{Introduction}
%---------------------------------------------------------------------------------------------------------------------------------------------------------------------------------------------------------------------------------------
Cosmic inflation, a period of accelerated expansion of the early universe, provides an excellent solution to cosmological problems, including the horizon, flatness, and monopole problems~\cite{Guth:1980zm, Linde:1981mu, Albrecht:1982wi, Sato:1980yn}, and is a mechanism to generate the primordial seeds required to explain the observed large-scale structure in the universe~\cite{Mukhanov:1981xt, Hawking:1982cz, Starobinsky:1982ee, Guth:1982ec, Bardeen:1983qw, Kodama:1985bj}. The most recent cosmic microwave background (CMB) data show that successful inflation models should predict a nearly scalar invariant and quasi-adiabatic Gaussian primordial curvature power spectrum with the amplitude of $\mathcal{P}_S(k_\ast)=2.1\times10^{-9}$ at the pivot scale $k_\ast=0.05~\text{Mpc}^{-1}$ and the spectral tilt of $n_S = 0.965\pm0.004$~\cite{Ade:2013zuv}. Inflation models also predict that  the second-order  scalar perturbations during inflation can source the gravitational waves (GWs)~\cite{Starobinsky:1979ty, Allen:1987bk, Sahni:1990tx, Matarrese:1992rp, Noh:2004bc, Nakamura:2004rm, Ananda:2006af, Baumann:2007zm, Alabidi:2012ex}. Thus, GWs induced by the second-order scalar perturbations have attracted much attention in recent years~\cite{Kohri:2018awv, Cai:2018dig, Fu:2019vqc, Inomata:2018epa, Inomata:2019zqy, Domenech:2020kqm}. Such scalar-induced GWs can be sizable if the curvature perturbations are enhanced at scales smaller than the CMB scale~\cite{Ananda:2006af, Baumann:2007zm, Alabidi:2012ex}. 

The gravitational collapse of over-dense regions in the early universe can form black holes~\cite{Zeldovich:1967lct, Hawking:1971ei, Carr:1974nx, Carr:1975qj}. In other words, the curvature perturbations exceeding a critical value collapses to form a primordial black hole (PBH) in the radiation-dominated (RD) era. The PBH formation is, therefore, a process of the threshold. In order to produce PBHs in the RD era, the primordial power spectrum for scalar modes at small scales should be enhanced by a factor of about $10^7$ with respect to its value at the CMB scale~\cite{Carr:2009jm}. Although there is no evidence, it was suggested that PBHs could be a natural candidate for dark matter (DM)~\cite{Ivanov:1994pa, Frampton:2010sw, Belotsky:2014kca, Carr:2016drx, Inomata:2017okj, Carr:2020xqk}. PBHs with very light masses are anticipated to Hawking radiate energetically. If the evaporation of PBHs is halted at some point, there remain stable remnants with Planck mass and size, and such remnants of PBHs can also be a natural candidate for DM~\cite{Adler:2001vs,Chen:2002tu, Chen:2004ft, Scardigli:2010gm, Dalianis:2019asr}. The true nature of DM, however, remains unclear. Therefore, in this work, we focus on the idea of PBHs as DM, which has taken a new flight in recent years due to the absence of well-motivated particle candidates for DM and recent observations of GWs from the binary merger of black holes~\cite{Bird:2016dcv, Sasaki:2016jop}. PBHs, if they are massive enough to avoid the Hawking evaporation, are considered to provide a significant fraction, if not all, of mysterious DM that dominate cosmic structures in the universe today.  

The enhancement in the primordial curvature power spectrum is often associated with a local feature of inflaton potentials, particularly at small scales~\cite{Ezquiaga:2017fvi, Mishra:2019pzq, Cheong:2019vzl}. In other words, by specifying the form of inflaton potential, one can provoke a local feature like a peak in the curvature power spectrum on scales relevant for the production of PBHs and GWs. However, not every inflaton potentials induce such enhancement; hence a certain degree of fine-tuning is needed~\cite{Hertzberg:2017dkh}. On the other hand, non-standard models of inflation often introduce additional contributions to the primordial power spectra of scalar and tensor modes; therefore, they could, in principle, give rise to the production of PBHs and secondary GWs without adjusting the inflaton potential~\cite{Lu:2019sti, Yi:2020cut, Gao:2020tsa, Yi:2020kmq, Lin:2020goi}. Thus, to investigate the enhancement mechanism in the curvature perturbations, we employ a single field inflationary model proposed in Ref~\cite{Tumurtushaa:2019bmc, Bayarsaikhan:2020jww}. The model acknowledges the effects of the derivative self-interaction of the scalar field and the kinetic coupling between the scalar field and gravity during inflation. However, it is challenging to achieve the correct values for the inflationary observable quantities, including the spectral indices and tensor-to-scalar ratio, on the CMB scale while enhancing the power spectrum on small scales such that the PBHs and GWs can be produced in the RD era. Still, this will be the focus of our work in this paper. The model we study is a subset of the so-called Horndeski theory of gravity~\cite{Horndeski:1974wa}, the most general four-dimensional scalar-tensor theory with equations of motion up to second order in derivatives of the scalar field, and its modern generalized Galileon formulation has been developed in Ref.~\cite{Deffayet:2011gz, Kobayashi:2011nu, Gao:2011vs, Gao:2012ib, Deffayet:2013lga}. 

The organization of the paper is as follows. In Sec.~\ref{sec:setup}, we review the single-field potential-driven inflationary models with the derivative self-coupling of the scalar field and the kinetic coupling between the scalar field and gravity. In Sec.~\ref{sec:powerspec}, we discuss the enhancement mechanism of primordial curvature perturbations for our model and present the formulae for the PBHs abundance as DM in Sec.~\ref{sec:PBHs} and the energy spectrum of GWs in Sec.~\ref{sec:GWs} produced in the RD era. In Sec.~\ref{sec:pot}, we reconstruct the inflaton potential and self-coupling functions from the power spectrum and its spectral tilt. Finally, Sec.~\ref{sec:conclusion} is devoted to conclusion and discussion. 
%---------------------------------------------------------------------------------------------------------------------------------------------------------------------------------------------------------------------------------------
\section{Equations of motion and Inflationary predictions}\label{sec:setup}
%---------------------------------------------------------------------------------------------------------------------------------------------------------------------------------------------------------------------------------------
The action of the cosmological models we investigate in this work is given as~\cite{Tumurtushaa:2019bmc, Bayarsaikhan:2020jww}
\begin{align}\label{eq:NMDCandGinf}
S = \int d^{4}x\sqrt{-g} \left[ \frac{M_{pl}^2}{2} R - \frac{1}{2} \left(g^{\mu\nu}-\frac{\alpha}{M^3}\xi(\phi)g^{\mu\nu}\partial_\rho\partial^\rho\phi +\frac{\beta}{M^2}G^{\mu\nu} \right)\partial_\mu\phi\partial_\nu \phi - V(\phi)\right]\,,
\end{align}
where the $\alpha$ and $\beta$ are dimensionless constants. Thus, by rescaling the $\alpha$ and $\beta$ to replace the mass scale $M$ with the reduced Planck mass $M_{pl}$, one can decrease the number of free parameters. The second term inside the parentheses reflects the derivative self-interaction of the scalar field $\phi$, while the third term indicates the kinetic coupling between the scalar field and gravity. $V(\phi)$ and $\xi(\phi)$ are the potential and self-coupling functions of the $\phi$, respectively.

Varying Eq.~(\ref{eq:NMDCandGinf})  with respect to spacetime metric $g_{\mu\nu}$, we obtain the Einstein equation 
\begin{eqnarray}\label{eq:Einsteineq}
&& G_{\mu\nu}=  M_{pl}^{-2}%T^{m ,r}_{\mu\nu}+ 
T_{\mu\nu}^{\phi}\,,
\end{eqnarray}
where the energy-momentum tensor for the scalar field is given by
\begin{align}
T_{\mu\nu}^{\phi} &=\partial_\mu\phi \partial_\nu\phi-\frac{1}{2}g_{\mu\nu}\left(\partial_\alpha\phi\partial^\alpha\phi+2V\right)\,\nonumber\\
&+\frac{\alpha}{2M^3}\left[( \xi \nabla_\alpha\phi\nabla^\alpha\phi)_{(\mu}\nabla_{\nu)}\phi - \xi\square\phi \nabla_\mu\phi\nabla_\nu\phi-\frac{1}{2}g_{\mu\nu}(\xi\nabla_\alpha\phi\nabla^\alpha\phi)_{\beta}\nabla^\beta\phi\right] \,\\
&+\frac{\beta}{M^2}\left[-\frac{1}{2}\nabla_\mu\phi \nabla_\nu\phi R + 2\nabla_\alpha\phi \nabla_{(\mu}\phi R^{\alpha}\,_{\nu)}+\nabla^\alpha\phi\nabla^\beta\phi R_{\mu\alpha\nu\beta}+\nabla_\mu\nabla^\alpha\phi\nabla_\nu\nabla_\alpha\phi\right.\nonumber\\
&\left.-\nabla_\mu\nabla_\nu\phi\square\phi-\frac12 G_{\mu\nu} \nabla_\alpha\phi \nabla^\alpha\phi+g_{\mu\nu}\left(-\frac{1}{2} \nabla^\alpha\nabla^\beta\phi\nabla_\alpha\nabla_\beta\phi+\frac{1}{2}\left(\square\phi\right)^2-\nabla_\alpha\phi\nabla_\beta\phi R^{\alpha\beta}\right)\right]\,.\nonumber
\end{align}
Using the Bianchi identity $\nabla^\mu G_{\mu\nu}=0$ in Eq.~(\ref{eq:Einsteineq}), we obtain the evolution equation for the scalar field as
\begin{eqnarray}
\nabla^{\mu}\,T_{\mu\nu}^\phi=0\,.
\end{eqnarray}
In a spatially flat Friedman-Robertson-Walker universe with metric 
\begin{eqnarray}\label{eq:flatMetric}
ds^2=-dt^2+a(t)^2\delta_{ij}dx^i dx^j\,,
\end{eqnarray} 
where $a(t)$ is a scale factor, the background equations of motion are obtained as
{\small \begin{align}
&3M_{pl}^2 H^2 = \frac{1}{2} \dot{\phi}^2 +V +\frac{3\alpha}{M^3}H\xi\dot{\phi}^3 \left( 1-\frac{\dot{\xi}}{6H\xi}\right) - \frac{9\beta}{2M^2}\dot{\phi}^2 H^2  \,,\label{eq:EE00}\\
& M_{pl}^2\left(2\dot{H}+3H^2\right) = -\frac{1}{2}\dot{\phi}^2 + V  -\frac{\alpha}{M^3}\xi\dot{\phi}^3\left(\frac{\ddot{\phi}}{\dot{\phi}} + \frac{\dot{\xi}}{2\xi }\right) +\frac{\beta\dot{\phi}^2}{2M^2}\left(2\dot{H} +3H^2 +4H\frac{\ddot{\phi}}{\dot{\phi}} \right)\,,\label{eq:EEii}\\
& \ddot{\phi} +3H\dot{\phi} + V_{,\phi} -\frac{\alpha}{2M^3}\dot{\phi}\left[\ddot{\xi}\dot{\phi}+3\dot{\xi}\ddot{\phi} - 6\xi\dot{\phi} \left( \dot{H} + 3H^2 + 2 H \frac{\ddot{\phi}}{\dot{\phi}} \right) \right] -\frac{3\beta}{M^2} H \dot{\phi}\left( 2\dot{H} +3H^2 +H\frac{\ddot{\phi}}{\dot{\phi}}\right)=0\,,\label{eq:fieldEq}
\end{align}} where $V_{,\phi}\equiv dV/d\phi$ and the dot denotes the derivative with respect to time. 

In a slow-roll scenario, inflation occurs as the scalar (\emph{or} inflaton) field rolls from some initial $\phi_i$ value to another $\phi_e$ value. $\phi_i$ and $\phi_e$ denote the beginning and end of inflation, respectively. For inflation to be successful, the rolling process must happen very slowly; therefore, during inflation, the potential energy of the scalar field is much larger than its kinetic energy, $V\gg\dot{\phi}^2$, and the friction is much larger than its acceleration, $3H\dot{\phi}\gg\ddot{\phi}$. To reflect the slow-roll inflation, we define the following new parameters,
\begin{align}\label{eq:srparam}
\epsilon_1 \equiv - \frac{\dot{H}}{H^2}\,,\quad \epsilon_2 \equiv -\frac{\ddot{\phi}}{H\dot{\phi}}\,, \quad \epsilon_3 \equiv \frac{\xi_{,\phi}\dot{\phi}}{\xi H}\,,\quad \epsilon_4 \equiv \frac{\xi_{,\phi\phi}\dot{\phi}^4}{V}\,,\quad \epsilon_5 \equiv \frac{\dot{\phi}^2}{2M_{pl}^2H^2}\,,
\end{align}
and require $|\epsilon_i|\ll 1$, where $i=1, 2, 3, 4, 5$. Then, Eq.~(\ref{eq:fieldEq}) can be rewritten in terms of these newly defined parameters as
{\small \begin{align}
3H\dot{\phi} \left[ 1- \frac{\epsilon_2}{3} - \frac{3\beta H^2}{M^2}\left( 1- \frac{2\epsilon_1}{3} - \frac{\epsilon_2}{3} \right) + \frac{3\alpha \xi H \dot{\phi}}{M^3} \left( 1- \frac{\epsilon_1}{3} - \frac{2\epsilon_2}{3} + \frac{2\epsilon_2\epsilon_3}{9}\right) \right] = -V_{,\phi}\left( 1- \frac{\alpha V \epsilon_4}{2M^3V_{,\phi}}\right)\,.
\end{align}}
In the limit $(\alpha, \beta) \rightarrow0$ during inflation, the above equation reduces to $3H\dot{\phi}\simeq -V_{,\phi}$. 

In addition to Eq.~(\ref{eq:srparam}), we introduce the following relation between the second and third terms inside the parenthesis in Eq.~(\ref{eq:NMDCandGinf}), 
\begin{align}\label{eq:gammaDef}
\gamma\equiv \frac{\alpha \xi g^{\mu\nu} \square \phi/M^3}{\beta G^{\mu\nu}/M^2} \,,
\end{align}
which weighs the contributions of each term. In other words, if the kinetic coupling between the scalar field and gravity is much stronger (or weaker) than the derivative self-interaction of the scalar field, then we get $\gamma \ll 1$ ($\gamma \gg 1$). Consequently, the $\gamma\sim\mathcal{O}(1)$ if both the second and third terms contribute equally during inflation. However, these terms cancel each other if the $\gamma=1$ because of their opposite signs in Eq.~(\ref{eq:NMDCandGinf}), and results, in that case, should be that of the standard slow-roll inflation. We are interested in the $\gamma\sim \mathcal{O}(1)$ limit in this work. 
Eq.~(\ref{eq:gammaDef}) is rewritten as
\begin{align}\label{eq:gammadef}
\gamma = \frac{\alpha}{\beta M}\frac{\xi \dot{\phi}}{H}\,,
\end{align}
and its implication is still the same. The importance of introducing the $\gamma\sim \mathcal{O}(1)$ parameter is to determine the shape of $\xi(\phi)$ using the observational constraints, which we will discuss in Sec.~\ref{sec:pot}.  Requiring the slow-roll conditions ($V\gg\dot{\phi}^2$ and $3H\dot{\phi}\gg\ddot{\phi}$) to be satisfied during inflation, we rewrite Eqs.~(\ref{eq:EE00}) and (\ref{eq:EEii}) as 
\begin{align}
3 M_{pl}^2 H^2 \simeq V\,,\qquad 3H\dot{\phi} \left( 1 + \mathcal{A}\right) \simeq -V_{,\phi}\,,\label{eq:sreq}
\end{align}
where 
\begin{align}\label{eq:defA}
\mathcal{A}\equiv \frac{3\alpha}{M^3}\xi \dot{\phi} H - \frac{3\beta}{M^2}H^2.
\end{align}
The equations in Einstein gravity are recovered for $|\mathcal{A}|\ll1$ in Eqs.~(\ref{eq:sreq}), and the limit $|\mathcal{A}|\gg1$ is known as the high friction limit~\cite{Germani:2010gm, Germani:2010ux, Tsujikawa:2012mk, Yang:2015zgh, Yang:2015pga, Sato:2017qau}. Deviation from Einstein's gravity is, therefore, reflected in $|\mathcal{A}|\sim 1$. Using Eq.~(\ref{eq:sreq}), we rewrite Eq.~(\ref{eq:srparam}) in the following form,
\begin{align}
\epsilon_1 \simeq \frac{\epsilon_V}{1+\mathcal{A}}\,, \quad \epsilon_2 \simeq \frac{\eta_V - 3\epsilon_V}{1+\mathcal{A}} + \frac{2\epsilon_V}{(1+\mathcal{A})^2}\,, \quad \epsilon_3 \simeq \frac{\eta_V - 4\epsilon_V}{1+\mathcal{A}} + \frac{2\epsilon_V }{(1+\mathcal{A})^2}\,,
\end{align}
where 
\begin{align}\label{eq:potSRparam}
\epsilon_V \equiv \frac{M_{pl}^2}{2}\left( \frac{V_{,\phi}}{V}\right)^2\,, \qquad \eta_{V} \equiv M_{pl}^2 \frac{V_{,\phi\phi}}{V}\,.
\end{align}
Inflation ends when the slope of the potential gets steep enough such that the first slow-roll condition is violated. Thus, the field value at the end of inflation is estimated by solving $\epsilon_1(\phi_e)=1$. 
The amount of inflation is quantified by the number of $e$-folds expressed as
\begin{align}\label{eq:efold}
N = \int^{\phi_{e}}_{\phi_i} \frac{H}{\dot{\phi}}d\tilde{\phi} \simeq \frac{1}{M_{pl}^2}  \int^{\phi_{e}}_{\phi_i} \frac{V}{V_{,\tilde{\phi}}}(1+\mathcal{A})d\tilde{\phi}\,.
\end{align}

%======================================================================Perturbation=============================================================================================================
The kinetic coupling between the scalar field and gravity and the inflaton self-interaction affect the dynamical evolution of the universe both at the background and perturbation level. Thus, their presence is imprinted in the observed power spectra of the scalar and tensor modes. Following~\cite{Kobayashi:2011nu, Gao:2011vs, Gao:2012ib, Tumurtushaa:2019bmc}, we perform the perturbation analyses of both the scalar and tensor modes in the uniform field gauge, where $\delta\phi(t,\vec{x})=0$. 

Let us begin with the perturbations of tensor modes characterized by the tensor part of the metric perturbations $h_{ij}$. Then, the quadratic action for the tensor perturbations reads~\cite{Kobayashi:2011nu, Gao:2011vs, Gao:2012ib, Tumurtushaa:2019bmc}
\begin{align}
S_T^{2} = \frac{M_{pl}^2}{8} \int dt d^3x a^3 \left[ G_T \dot{h_{ij}}^2 - \frac{F_T}{a^2} \left( \partial_k h_{ij} \right)^2 \right]\,,
\end{align}
where 
\begin{align}
G_T = 1 -  \frac{\beta}{2M^2M_{pl}^2}\dot{\phi}^2\,,\qquad F_T = 1+ \frac{\beta}{2M^2M_{pl}^2}\dot{\phi}^2\,.
\end{align}
%======================================================================footnote================================================================================================================
%~\footnote{which can also be written in terms of the slow-roll parameters as
%\begin{align}
%G_T = 1 +  \frac{\epsilon_V}{3(1-\gamma)(1+\mathcal{A})}\,,\qquad F_T = 1 -   \frac{\epsilon_V}{3(1-\gamma)(1+\mathcal{A})}\,,
%\end{align}
%where we used 
%\begin{align}
%\mathcal{A} = -\frac{3\beta}{M^2}(1-\gamma) &\implies \frac{\beta}{M^2} = -\frac{\mathcal{A}}{3H^2(1-\gamma)}\,, \\
%3 H \dot{\phi} (1+\mathcal{A}) \simeq -V_{,\phi} &\implies \frac{\dot{\phi}^2}{2M_{pl}^2} = \frac{H^2}{(1+\mathcal{A})}\epsilon_1\,,\\
%G_T = 1-\frac{\beta}{2M^2 M_{pl}^2} \dot{\phi}^2 &\implies G_T = 1 + \frac{\mathcal{A}}{3(1-\gamma)(1+\mathcal{A})} \epsilon_1 \\
%& \qquad\qquad\, = 1+ \frac{\epsilon_V}{3(1-\gamma)(1+\mathcal{A})} - \frac{\epsilon_V}{3(1-\gamma)(1+\mathcal{A})^2}.
%\end{align} }
%============================================================================================================================================================================================
The evolution equation of the tensor perturbation modes is written as
\begin{align}
u_{\lambda, k}'' + \left( c_T^2 k^2 - \frac{\mu_T^2 -1/4}{\tau^2} \right)u_{\lambda, k} = 0\,,
\end{align}
where $u_{\lambda, k} \equiv z_T h_{\lambda, k}$ with $z_T=(a/2) \left( G_T F_T \right)^{1/4}$, $c_T^2\equiv F_T/G_T$, and $\mu_T\simeq 3/2 + \epsilon_1$. Here, $\tau$ is conformal time and $\lambda= \times, +$ denotes the two polarization modes of the tensor perturbations.
%======================================================================footnote================================================================================================================
%~\footnote{ are given as
%\begin{align}
%c_T^2 = 1 -  \frac{2 \epsilon_V}{3(1-\gamma)(1+\mathcal{A})} \,,\quad \mu_T \simeq \frac{3}{2} + \frac{\epsilon_V}{1+\mathcal{A}}\,.
%\end{align}}
%============================================================================================================================================================================================
If one assumes the Bunch-Davies vacuum for the initial fluctuation modes at $c_T k |\tau| \gg 1$ and the constant slow-roll parameters, the solution to the above equation is
\begin{align}
u_{\lambda, k} = 2^{\mu_T-\frac{3}{2}} \frac{\Gamma(\mu_T)}{\Gamma(3/2)}\frac{(-c_T k \tau)^{\frac{1}{2}-\mu_T}}{\sqrt{2c_Tk}} e^{i \left( \mu_T -\frac{1}{2}\right)\frac{\pi}{2} }\,.
\end{align}
The power spectrum of the tensor modes on the large scale, $c_T k |\tau| \ll 1$, is obtained as
\begin{align}\label{eq:tensorPt}
\mathcal{P}_T =\frac{k^3}{\pi^2} \sum_{\lambda} \left| u_{\lambda, k}{z_T} \right|^2 \simeq \frac{H^2}{2\pi^2 M_{pl}^2 c_T^3}\,.
\end{align}
The spectral tilt of the tensor power spectrum is computed at the time of horizon crossing as
\begin{align}
n_T \equiv \left. \frac{d\ln \mathcal{P}_T}{d\ln k} \right|_{c_T k = a H} = 3 - 2\mu_T \simeq -\frac{2\epsilon_V}{1+\mathcal{A}}\,.
\end{align}
Now, let us consider the perturbations for scalar modes $\zeta$. The quadratic action for the scalar perturbation is written as~\cite{Kobayashi:2011nu, Gao:2011vs, Gao:2012ib, Tumurtushaa:2019bmc}
\begin{align}
S_S^{2} = M_{pl}^2 \int dt d^3x a^3 \left[ G_S \dot{\zeta}^2 - \frac{F_S}{a^2} \left( \partial_i \zeta \right)^2 \right]\,,
\end{align}
where 
\begin{align}
G_S = \frac{\Sigma}{\Delta^2} G_T^2 + 3G_T\,, \quad F_S = \frac{1}{a}\frac{d}{dt}\left( \frac{a}{\Delta} G_T^2\right) - F_T\,,
\end{align} 
with 
\begin{align}
&\Sigma = \frac{1}{2}\dot{\phi}^2  - 3M_{pl}^2 H^2 + \frac{9\beta}{M^2} H^2 \dot{\phi}^2 + \frac{\alpha}{M^3} \xi \dot{\phi}^3 H \left(6 - \frac{\dot{\xi}}{\xi H}\right)\,, \nonumber\\
&\Delta = M_{pl}^2 H \left( 1 - \frac{3\beta}{2M^2M_{pl}^2}\dot{\phi}^2\right) - \frac{\alpha}{2M^3}\xi\dot{\phi}^2\,.
\end{align}
The perturbation equations for the scalar modes reads
\begin{align}\label{eq:master}
v_{k}'' + \left( c_S^2 - \frac{\mu_S^2 -1/4}{\tau^2} \right) v_{k} =0\,,
\end{align}
where $v_{k} \equiv z_S \zeta_k$ with $z_S=\sqrt{2}a(G_S F_S)^{1/4}$, $c_S^2\equiv F_S/G_S$, and $\mu_S\simeq 3/2 +\epsilon_1 -\epsilon_2$.
%======================================================================footnote================================================================================================================
%~\footnote{ are given as 
%\begin{align}
%c_S^2 = 1 + \left(\frac{3\gamma^2-15\gamma -14}{1+\gamma}\right) \frac{\epsilon_V}{1+\mathcal{A}}\,, \quad \mu_S \simeq \frac{3}{2} + \frac{4\epsilon_V - \eta_V}{1+\mathcal{A}} -\frac{\epsilon_V}{(1+\mathcal{A})^2}\,.
%\end{align}}
%============================================================================================================================================================================================
Taking the Bunch-Davies vacuum for the initial fluctuations into account, we obtain the solution to the scalar perturbation equation as 
\begin{align}
v_{k} = 2^{\mu_S -\frac{3}{2}} \frac{\Gamma(\mu_S)}{\Gamma(3/2)}\frac{(-c_S k \tau)^{\frac{1}{2}-\mu_S}}{\sqrt{2 c_S k}} e^{i\left( \mu_S -\frac{1}{2}\right)\frac{\pi}{2}}\,.
\end{align}
The power spectrum of the scalar perturbations on large scales, $c_T k |\tau| \ll 1$, is computed as 
\begin{align}\label{eq:scalarPs}
\mathcal{P}_S = \frac{k^3}{2\pi^2}\left| \frac{v_k}{z_S}\right|^2 &\simeq \frac{H^2}{8\pi^2M_{pl}^2 c_S^2 \epsilon_V}(1+\mathcal{A}) =  \frac{V^3}{12\pi^2 M_{pl}^6 V_{,\phi}^2} (1+\mathcal{A})\,,
\end{align}
where $\mathcal{A}$ is given in Eq.~(\ref{eq:defA}). %, and the standard result in Einstein gravity with the scalar field is obtained in the $\mathcal{A}\ll 1$ limit. 
An implication of Eq.~(\ref{eq:scalarPs}) is that the scalar power spectrum can be enhanced if the function $\mathcal{A}$ is large enough, $\mathcal{A}\gg 1$. Therefore, if the power spectrum is enhanced on scales smaller than the CMB scale due to the presence of $\mathcal{A}$,  %\sout{ function $\mathcal{A}$can, in principle, enhance the scalar power spectrum on scales smaller than the CMB scale; hence} 
(i.) the PBHs can be formed, and (ii.) the induced GWs can also be generated. We devote the following sections to such possibilities. 

The spectral tilt of the curvature perturbations at the time of horizon crossing gets
\begin{align}\label{eq:nstilt}
n_S-1 = \left. \frac{d\ln \mathcal{P}_S}{d\ln k} \right|_{c_Sk = aH} =3-2\mu_S =\frac{2}{1+\mathcal{A}} \left[\eta_V -  \frac{3+4\mathcal{A}}{1+\mathcal{A}} \epsilon_V \right]\,,
\end{align}
and the tensor-to-scalar ratio reads 
\begin{align}\label{eq:ttos}
r\equiv \frac{\mathcal{P}_T}{\mathcal{P}_S} \simeq \frac{16\epsilon_V}{1+\mathcal{A}}\,.
\end{align}
As we mentioned earlier that the $|\mathcal{A}|\ll 1$ limit is equivalent to Einstein gravity. Thus, the above results reduce to that of the standard single-filed inflation in Einstein gravity: $n_S-1=-6\epsilon_{V}+2\eta_{V}$ and $r=16\epsilon_{V}$. The deviation from GR is implied for the opposite $|\mathcal{A}|\gg 1$ case, and the predictions on $n_S$, $n_T$, and $r$ are suppressed by a factor of $\mathcal{A}$: $n_S-1=(-8\epsilon_{V}+2\eta_{V})/\mathcal{A}$, $n_T=-2\epsilon_V/\mathcal{A}$, and $r=16\epsilon_{V}/\mathcal{A}$~\cite{Tumurtushaa:2019bmc}.
%======================================================================Perturbation end here =====================================================================================================

%---------------------------------------------------------------------------------------------------------------------------------------------------------------------------------------------------------------------------------------
\section{The production of PBH and GW}\label{sec:powerspec}
%---------------------------------------------------------------------------------------------------------------------------------------------------------------------------------------------------------------------------------------

Eq.~(\ref{eq:scalarPs}) shows that the power spectrum of curvature perturbations can be enhanced if the function $\mathcal{A}$ is large enough. On the other hand, sufficiently large curvature fluctuations can form PBHs due to the gravitational collapse when they reenter the horizon in the RD era~\cite{Zeldovich:1967lct, Hawking:1971ei, Carr:1974nx, Carr:1975qj}, which consequently implies that the formation of PBH may be possible for our model. Since the CMB temperature and polarization measurements constrain the primordial perturbations to be very small at large scales, we are interested in enhancing the curvature power spectrum on scales smaller than the CMB scale, \emph{i.e.,} $k_\text{PBH} \gg k_\ast$. Besides, the sufficiently large density fluctuations generated during inflation can also simultaneously produce a substantial amount of GWs in the RD era~\cite{Ananda:2006af, Baumann:2007zm}. We investigate such possibilities of the PBH and GW productions in this section.

\begin{figure}[h!]
\centering  
\includegraphics[width=0.6\textwidth]{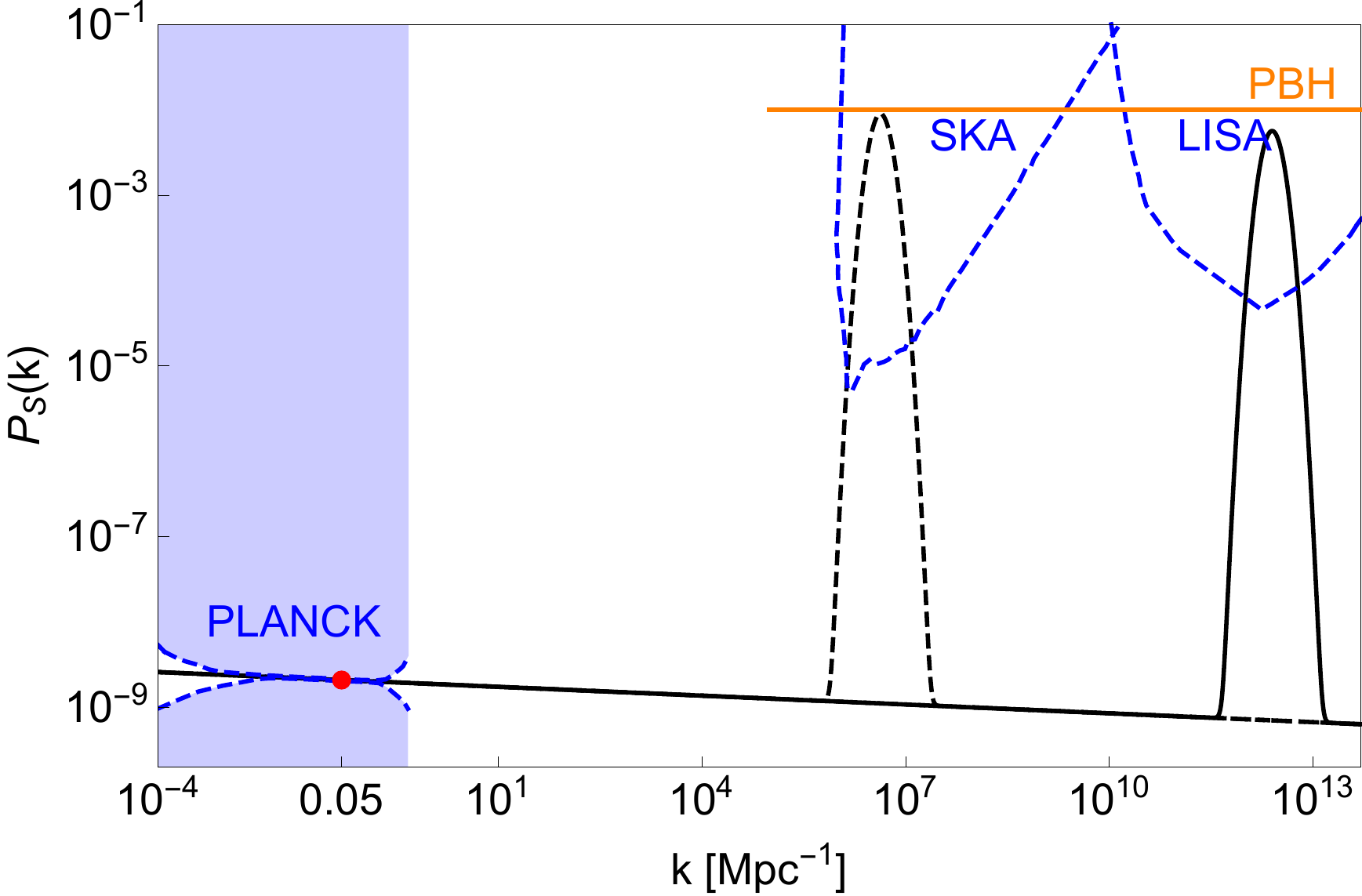} \qquad 
\caption{The primordial scalar power spectrum as a function of wavenumber $k$. The peaks with an amplitude $A_S= 6\times 10^6$ and a width $\sigma=0.3$ occur at $k_\text{PBH}=4.3\times10^{6}\,\text{Mpc}^{-1}$ and $k_\text{PBH}=2.5\times10^{12}\,\text{Mpc}^{-1}$, dashed and solid black lines, respectively. The red point at $k_\ast=0.05~\text{Mpc}^{-1}$ denotes the CMB spectrum $\mathcal{P}_S(k_\ast)=2.1\times10^{-9}$ with the spectral tilt $n_S=0.9655$. The dashed blue lines are the CMB constraints on the $\mathcal{P}_S(k)$~\cite{Ade:2013zuv} and GW limits from SKA and LISA, where the shaded regions are excluded~\cite{Inomata:2018epa}. The orange line at $\mathcal{P}_S(k_\text{PBH})\sim 10^{-2}$ indicates the required amplitude of the power spectra to form PBHs~\cite{Green:2020jor}.}\label{fig:powerspectrum}
\end{figure}

The enhancement mechanism of the power spectrum on small scales is often associated with a local feature of the inflaton potential%, \emph{i.e.,} a local feature of inflaton potential plays an important role both in inflation and in the RD era
~\cite{Ezquiaga:2017fvi, Mishra:2019pzq}. However, as we mentioned in the Introduction, not every inflaton potentials induce such enhancement, which is a common drawback in these approaches. Therefore, to investigate the formation of PBHs and GWs, we first specify the primordial power spectrum instead of the inflaton potential. Then, construct the inflaton potential from the power spectrum. We propose the following parameterization of the primordial power spectrum~\cite{Cai:2018dig, Inomata:2018epa, Kohri:2020qqd}
\begin{align}\label{eq:powerSs}
\mathcal{P}_S(k) = \mathcal{P}_S(k_\ast) \left( \frac{k}{k_\ast} \right)^{n_S-1}\left[ 1 + \frac{A_S}{\sqrt{2\pi \sigma^2}} e^{-\frac{1}{2\sigma^2} \left(\ln \frac{k}{k_\text{PBH}} \right)^2 }\right]\,,
\end{align}
where $\mathcal{P}_S(k_\ast)$ is the amplitude of the CMB spectrum at the pivot scale $k_\ast$ and $n_S$ is the spectral tilt given in Eq.~(\ref{eq:nstilt}). 
%%========================From k to N ========================================================================================================================================================
%~\footnote{\footnotesize{Since $H\simeq\text{const.}$ during inflation, one can write in Eq.~(\ref{eq:AasN}): $$\ln\left( \frac{k}{k_\text{PBH}}\right)=\ln\left( \frac{k_\ast}{k_\text{PBH}}\frac{k}{k_\ast}\right)=\ln\left[ \frac{(a H)_\ast}{(a H)_\text{PBH}} \frac{(a H)_k}{(a H)_\ast}\right]=\ln\left(\frac{a_\ast}{a_\text{PBH}}\right)+\ln\left(\frac{a_k}{a_\ast}\right)=N_k-(N_\ast-N_\text{PBH})=N_k-N_p.$$ Here $k_\ast \lesssim k \lesssim k_\text{max}$ and $k_\ast \lesssim k_\text{PBH} \lesssim k_\text{max}$, therefore:
%\begin{itemize}
%\item If $k=k_\ast$, then $N_k=0$ hence $\bar{N}=-(N_\ast-N_\text{PBH}) =-N_p$. 
%\item If $k=k_\text{PBH}$, then $N_k=N_\ast-N_\text{PBH}$ hence $\bar{N}=0$. 
%\item If $k=k_\text{max}$, then $N_k=N_\ast$ hence $\bar{N}=N_\text{PBH}$.
%\end{itemize}%The range of $N_k$ is given by $0\leq N_k\leq N_\ast$. }}
%%========================================================================================================================================================================================= 
The $A_S$ and $\sigma$ are free parameters determining the height and width of the Gaussian peak, respectively. %~\footnote{The peak is called narrow if $\sigma\ll1$ and broad if  $\sigma\gtrsim\mathcal{O}(1)$~\cite{Cai:2018dig}.  } 
To form PBHs on scales smaller than the CMB scale ($k_\text{PBH}\gg k_\ast$), the $\mathcal{P}_S(k_\text{PBH})$ should be enhanced by a factor of about $10^7$ with respect to its value at the CMB scale, which is possible mainly due to the peak amplitude $A_S/\sqrt{2\pi\sigma^2}$. In Fig.~\ref{fig:powerspectrum}, we plot the $k$ dependence of Eq.~(\ref{eq:powerSs}) in light of the several observational limits and constraints adopted from Ref.~\cite{Inomata:2018epa}.  In particular, the orange line at $\mathcal{P}_S(k_\text{PBH})\sim 10^{-2}$ indicates the required amplitude of power spectra for the PBHs formation~\cite{Green:2020jor}. The figure also implies that the power spectra that have peaks at either $k\sim \mathcal{O}(10^{6})\, \text{Mpc}^{-1}$ or $k\sim \mathcal{O}(10^{12})\, \text{Mpc}^{-1}$ lead to the production of stochastic GW background in the frequency band targeted by the pulsar timing array (PTA) experiments and the space-based interferometers~\cite{Inomata:2018epa}. 
Then, to explain the small scale enhancement as parameterized in Eq.~(\ref{eq:powerSs}), we will reconstruct the inflaton potential and the self-coupling functions  in Sec.~\ref{sec:pot} from the power spectrum in Eq.~(\ref{eq:powerSs}) and its spectral tilt in Eq.~(\ref{eq:nstilt}).

%---------------------------------------------------------------------------------------------------------------------------------------------------------------------------------------------------------------------------------------
\subsection{Primordial black holes as dark matter}\label{sec:PBHs}
%---------------------------------------------------------------------------------------------------------------------------------------------------------------------------------------------------------------------------------------
When the curvature perturbations that left the Hubble horizon during inflation reenter the horizon in the RD era, they generate over-dense regions in the universe. The over-density would grow as the universe expands and eventually exceeds a critical value when the comoving size of such region becomes the order of the horizon size. Consequently, all the fluctuations inside the Hubble volume would immediately collapse to form PBHs~\cite{Zeldovich:1967lct, Hawking:1971ei, Carr:1974nx, Carr:1975qj}. The mass of the formed PBHs is assumed to be proportional to the mass inside the Hubble volume, \emph{i.e.,}  the horizon mass, at that time~\cite{Ozsoy:2018flq}%$M_\text{horizon} = (2GH)^{-1}$,
\begin{align}
M_{PBH} &= \tilde{\gamma} M_\text{horizon} = 2.43 \times 10^{12} \left(\frac{\tilde{\gamma}}{0.2}\right)\left(\frac{g_\ast}{106.75}\right)^{-\frac{1}{6}}\left(\frac{k}{\text{Mpc}^{-1}}\right)^{-2} M_\odot\,,
\end{align}
where the $M_\odot$ and $\tilde{\gamma}$ denote  the solar mass and efficiency of gravitational collapse, respectively.  Although the value of  $\tilde{\gamma}$ depends on the details of gravitational collapse, it has a typical value of $\tilde{\gamma}=0.2$~\cite{Carr:1975qj}, which we adopt in this study. 

The standard treatment of PBH formation is based on the Press-Schechter model of gravitational collapse~\cite{Press:1973iz}. The energy density fraction of PBHs of mass $M_\text{PBH}$ at the formation to the total mass of the universe is denoted by $\beta(M_\text{PBH})\equiv \rho_\text{PBH}/\rho$, where $\rho$ is the energy density of the universe and $\rho_{PBH}$ is the energy density of PBHs. The PBHs would be formed if the density contrast $\delta\equiv \delta\rho/\rho$ of the over-dense regions dominates a certain threshold $\delta_c$ value~\cite{Green:2004wb, Harada:2013epa, Musco:2018rwt}.
The distribution function of the density perturbation is assumed to be governed by the Gaussian statistics. Thus, one can write the production rate of PBHs with mass $M(k)$ as follows~\cite{Ozsoy:2018flq}%~\cite{Green:2004wb, Harada:2013epa, Musco:2018rwt}
\begin{align}
\beta(M) &= \int_{\delta_c} \frac{d\delta}{\sqrt{2\pi\bar{\sigma}^2(M)}}e^{-\frac{\delta^2}{2\bar{\sigma}^2(M)}} 
=\sqrt{\frac{2\bar{\sigma}^2(M)}{\pi \delta_c^2}}e^{-\frac{\delta_c^2}{2\bar{\sigma}^2(M)}} \,,
\end{align}
where the variance $\bar{\sigma}^2(M)$ represents the coarse-grained density contrast with the smoothing scale $k$ and is defined as~\cite{Young:2014ana, Ozsoy:2018flq}
\begin{align}
\bar{\sigma}^2(M)  &= \int \frac{dq}{q} W^2(q k^{-1}) \mathcal{P}_\delta(q) = \frac{16}{81} \int d\ln q\, W^2(q k^{-1})\left(qk^{-1}\right)^4 \mathcal{P}_\mathcal{S}(q)\,, 
\end{align}
where $\mathcal{P}_S(k)$ and $\mathcal{P}_\mathcal{\delta}(k)$ denote the dimensionless power spectra of the primordial comoving curvature perturbations and density perturbations, respectively. For the window function  $W(x)$, we adopt the Gaussian function $W(x) = e^{-x^2/2}$. The current fraction of PBH abundance in total DM today can be determined by~\cite{Ozsoy:2018flq}
\begin{align}
f_\text{PBH} \left( M\right) \equiv \frac{\Omega_\text{PBH}}{\Omega_\text{DM}} = \int \frac{dM}{M}f_\text{PBH} \left( M_\text{PBH}\right)\,,
\end{align}
where $\Omega_\text{DM}$ is the current density parameter of DM and 
\begin{align}\label{eq:pbh}
f_\text{PBH} \left( M_\text{PBH}\right) &= \frac{1}{\Omega_\text{DM}}\frac{d \Omega_\text{PBH}}{d\ln M_\text{PBH}}\nonumber\\
&\simeq 0.28 \times 10^8 \left( \frac{\gamma}{0.2}\right)^\frac32 \left(\frac{g_\ast}{106.75} \right)^{-\frac14} \left( \frac{\Omega_\text{DM}h^2}{0.12}\right)^{-1} \left( \frac{M_\text{PBH}}{M_\odot}\right)^{-\frac12} \beta(M_{\text{PBH}})\,,
\end{align}
where the effective number of relativistic degrees of freedom is $g_\ast = 106.75$ deep inside the RD era and $\Omega_\text{DM}h^2\simeq0.12$~\cite{Ade:2013zuv}. 

\begin{figure}[h!]
\centering 
\includegraphics[width=0.496\textwidth]{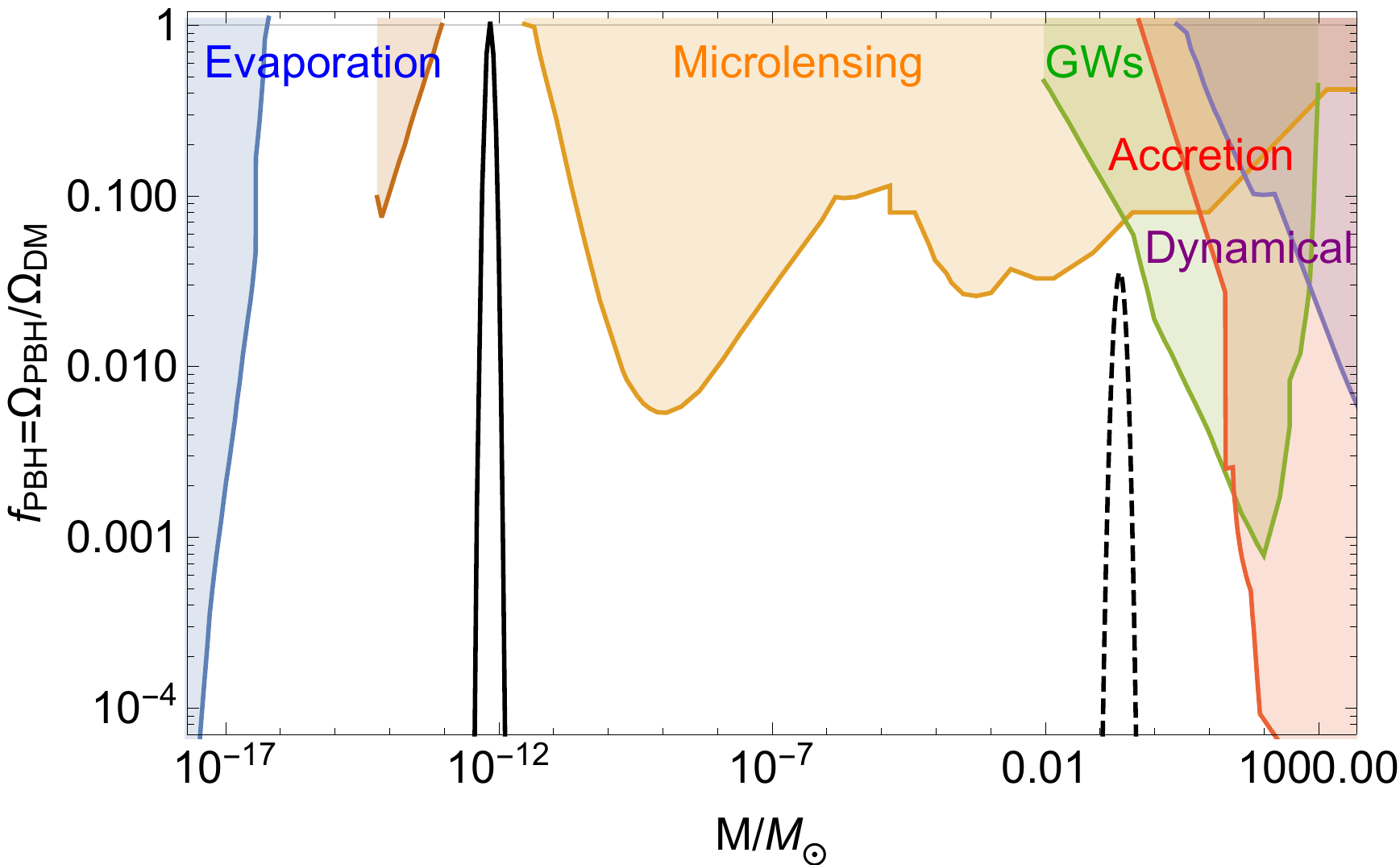} 
\includegraphics[width=0.496\textwidth]{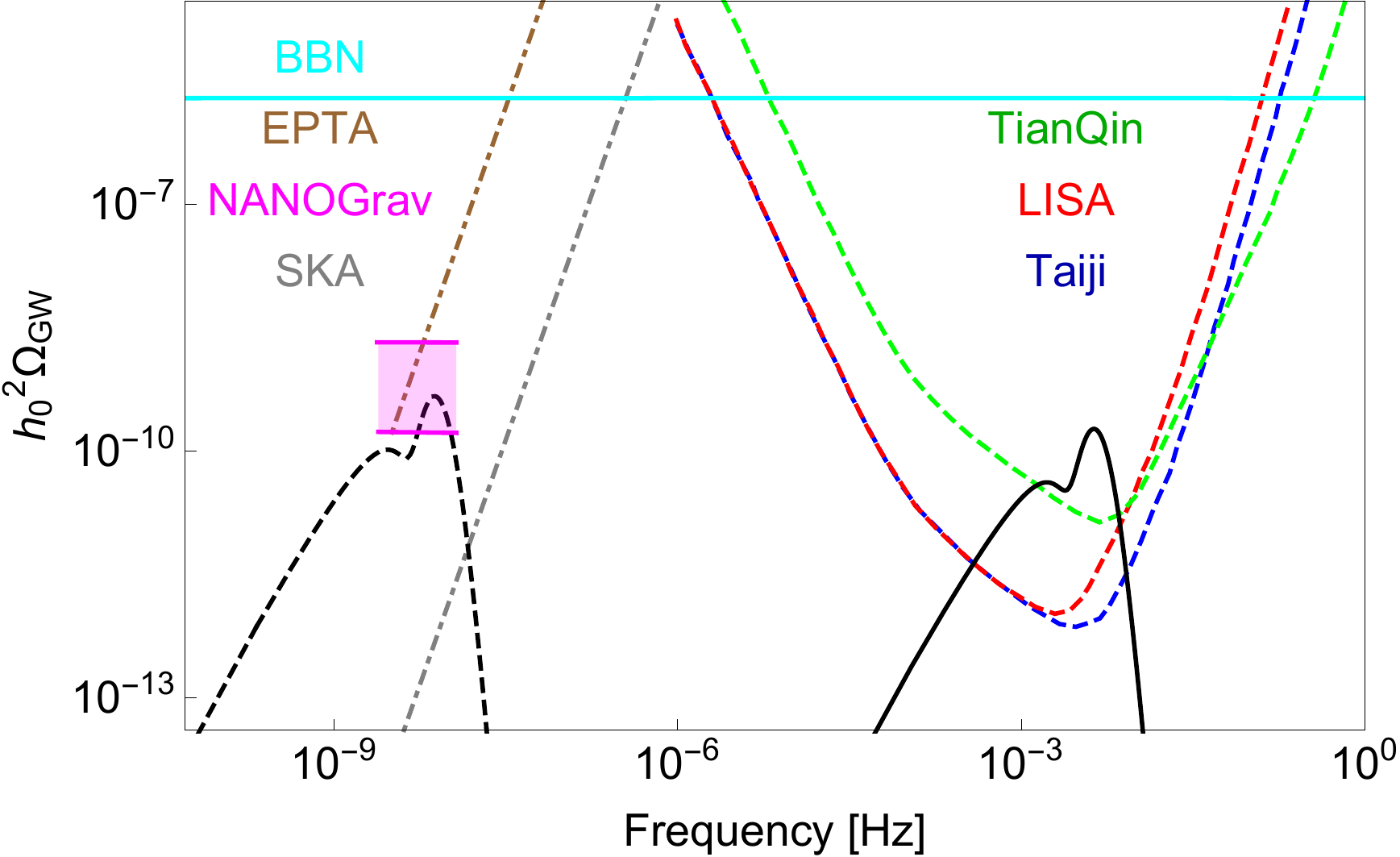}  \label{fig:gws}
\caption{\emph{Left}: The fraction of PBH abundance in total DM density. The shaded regions are excluded by the observational constraints, adapted from Ref.~\cite{Green:2020jor}. \emph{Right}: The energy spectra of scalar-induced GWs. The colored lines represent the sensitivity curves of the current and future GW projects~\cite{Audley:2017drz, Hu:2017mde,Luo:2015ght, Ferdman:2010xq, McLaughlin:2013ira, Moore:2014lga}. Numerical inputs are adopted from Fig.~\ref{fig:powerspectrum}. %The region above the solid cyan line is excluded by BBN observations.
} \label{fig:pbh}
\end{figure}

In the left panel of Fig.~\ref{fig:pbh}, we plot the current fraction of PBH abundance in total DM today as a function of mass $M$ from Eq.~(\ref{eq:pbh}). The figure also presents the currently available $f_\text{PBH}(M)$ constraints from different datasets over various mass ranges as summarized in Ref.~\cite{Green:2020jor}. The result shows that different peak scales in the scalar power spectrum lead to PBHs with different masses. The enhanced power spectra with the $k_\text{PBH}=2.45\times10^{12}\,\text{Mpc}^{-1}$ and $k_\text{PBH}=4.3\times10^{6}\,\text{Mpc}^{-1}$ correspond to PBHs with the peak mass around $10^{-12}M_\odot$ and $1M_\odot$, the solid and dashed black lines in the figure, respectively. The corresponding peak abundances are $f_\text{PBH}\sim1$ and $f_\text{PBH}\sim 0.05$. The peak abundance of PBHs depend on the critical density $\delta_c$, and it gets smaller as the $\delta_c$ value increases and vice versa. For PBHs with the peak mass of $M\sim10^{-12}M_\odot$ and abundances of $f_\text{PBH}\sim1$ to form, the critical density contrast should be $\delta_c\gtrsim0.145$. Thus, PBHs with the peak mass around $10^{-12}M_\odot$ can explain almost all the DM in the universe today. When the same $\delta_c$ value is applied, the PBHs with the peak mass around $1M_\odot$ can make up to $5\%$ of the observed DM abundance today. 
%---------------------------------------------------------------------------------------------------------------------------------------------------------------------------------------------------------------------------------------  abundance $f_\text{PBH}(M)$ with different masses
\subsection{Scalar-induced secondary gravitational waves}\label{sec:GWs}
%---------------------------------------------------------------------------------------------------------------------------------------------------------------------------------------------------------------------------------------

The sufficiently large density fluctuations generated during inflation can simultaneously produce a substantial amount of GWs when they reenter the horizon in the RD era~\cite{Ananda:2006af, Baumann:2007zm}. To investigate the GWs background evolving in the RD era, we assume negligible effects of inflaton field on cosmic evolution after reheating~\cite{Fu:2019vqc}. The equations of motion for the GW amplitude $h_{\bf{k}}(\tau)$ in Fourier space is written as~\cite{Ananda:2006af, Baumann:2007zm}
\begin{align}\label{eq:gweq}
h''_{\bf{k}} + 2\mathcal{H} h'_{\bf{k}}+k^2 h_{\bf{k}} =4 \mathcal{S}_{\bf{k}}\,,
\end{align}
where the prime denotes the derivative with respect to conformal time $\tau$. The conformal Hubble parameter $\mathcal{H}=a'/a$ is given in terms of the scale factor $a(\tau)=a_0(\tau/\tau_0)^{2/(1+3\omega)}$, where $\omega\equiv p/\rho$ is the equation-of-state parameter. The source term $\mathcal{S}_{\bf{k}}(\tau)$, which is a convolution of two first-order scalar perturbations at different wave numbers, is given by~\cite{Ananda:2006af, Baumann:2007zm}
\begin{align}\label{eq:source}
\mathcal{S}_{\bf{k}} = \int \frac{d^3\tilde{k}}{(2\pi)^{3/2}} \epsilon^{ij}({\bf{k}}) \tilde{k}_i \tilde{k}_j \left[ 2 \Phi_{\tilde{{\bf{k}}}} \Phi_{{\bf{k}}-\tilde{{\bf{k}}}} + \frac{4}{3(1+\omega)} \left(\frac{\Phi_{\tilde{{\bf{k}}}}'}{\mathcal{H}} + \Phi_{\tilde{{\bf{k}}}}\right) \left( \frac{\Phi_{{\bf{k}}-\tilde{{\bf{k}}}}'}{\mathcal{H}} + \Phi_{{\bf{k}}-\tilde{{\bf{k}}}} \right)\right]\,,
\end{align} 
where $\epsilon^{ij}({\bf{k}})$ is the polarization tensor. In the RD era, the scalar part of metric perturbations $\Phi_{{\bf{k}}}$ in Fourier space satisfies~\cite{Fu:2019vqc}
\begin{align}
\Phi_{{\bf{k}}}'' + \frac{4}{\tau} \Phi_{{\bf{k}}}'+ \frac{k^2}{3}\Phi_{{\bf{k}}} = 0\,,
\end{align}
which admits a solution~\cite{Baumann:2007zm}
\begin{align}
\Phi_{\bf{k}}(\tau)=\frac{9}{(k\tau)^2}\left[ \frac{\sin(k\tau/\sqrt{3})}{k\tau/\sqrt{3}} -\cos(k\tau/\sqrt{3}) \right]\zeta_{\bf{k}}\,,
\end{align}
where $\zeta_{{\bf{k}}}$ is the comoving curvature perturbations, giving rise to the curvature power spectrum of $\langle \zeta_{\bf{k}} \zeta_{\bf{\tilde{k}}} \rangle = (8\pi^2/9k^3) \mathcal{P}_S(k)\delta(\bf{k}+\bf{\tilde{k}})$. 
The power spectrum $\mathcal{P}_S(k)$ is given by Eq.~(\ref{eq:scalarPs}). Using the Green function method, one obtains the particular solutions for the Eq.~(\ref{eq:gweq}) as
\begin{align}
h_{\bf{k}} (\tau) = \int ^{\tau}_{\tau_0}G_{\bf{k}}(\tau,\tilde{\tau})\frac{a(\tilde{\tau})}{a(\tau)}\mathcal{S}_{\bf{k}}(\tilde{\tau})d\tilde{\tau}\,,
\end{align}
where the Green's function $G_{\bf{k}}(\tau, \tilde{\tau})$ obeys the equation of motion  
\begin{align}\label{eq:eqGreen}
G_{\bf{k}}''(\tau, \tilde{\tau})+\left(k^2 -\frac{a''}{a} \right)G_{\bf{k}}(\tau,\tilde{\tau})=\delta(\tau-\tilde{\tau})\,.
\end{align}
The exact solution to Eq.~(\ref{eq:eqGreen}) in the RD era is obtained to be $G_{\bf{k}}(\tau, \tilde{\tau})=\sin[k(\tau-\tilde{\tau})]/k$~\cite{Baumann:2007zm}.

The power spectrum of the induced GWs is then defined as 
\begin{align}\label{eq:hh}
\langle h_{\bf{k}}(\tau) h_{\tilde{{\bf{k}}}}(\tau)\rangle = \frac{2\pi^2}{k^3}\delta^{(3)}({\bf{k}}+\tilde{{\bf{k}}}) \mathcal{P}_T(k, \tau)\,,
\end{align}
and the fractional energy density per logarithmic wavenumber interval is 
\begin{align}\label{eq:GWenergy}
\Omega_{GW}(k,\tau) = \frac{1}{\rho_\text{tot}}\frac{d \rho_{GW}}{d \ln k}
=\frac{1}{24} \left( \frac{k}{a H}\right)^2\overline{\mathcal{P}_T(k,\tau)}\,,
\end{align}
where $a H=\mathcal{H}=2/[(3\omega+1)\tau]$, and the overline indicates the oscillation time average. As the relevant modes reenter the horizon, GWs are assumed to be induced instantaneously~\cite{Baumann:2007zm}. Thus, at the time of horizon reentry, we have $h_{\bf{k}}\sim \mathcal{S}_{\bf{k}}/k^2$; hence it gets contributions from all scalar modes $\Phi_{\tilde{k}}$. One can see from Eqs.(\ref{eq:hh})--(\ref{eq:GWenergy}) that $k^3\tilde{k}^3/|\bf{k}-\bf{\tilde{k}}|^3$ appears in the integrand in $\mathcal{P}_T$. The main contributions to $\mathcal{P}_T$ are from $\tilde{\bf{k}}$ that are closer to $\bf{k}$. Since the source $\mathcal{S}_{\bf{k}}$ decays as $a^{-p}$ with $3\leq p \leq 4$, the induced GWs propagate freely soon after horizon reentry and $h_{\bf{k}}\sim a^{-1}$. Thus, $\Omega_{GW}(k, \tau)\simeq \text{const.}$ well inside the horizon~\cite{Baumann:2007zm}.   

After combining Eqs.~(\ref{eq:source})--(\ref{eq:GWenergy}) and executing a straightforward yet tedious calculation, we obtain the power spectrum of the induced GWs in the RD era as~\cite{Ananda:2006af, Baumann:2007zm}
\begin{align}\label{eq:psSec}
\mathcal{P}_T(k,\tau) = 4\int^\infty_0 dv \int^{1+v}_{|1-v|} du \left[ \frac{4v^2-(1-u^2+v^2)}{4uv}\right]^2 I_{RD}^2(u,v,x) \mathcal{P}_S(kv)\mathcal{P}_S(ku)\,,
\end{align}    
where $u=|{\bf{k}-\bf{\tilde{k}}|}/k$, $v=\tilde{k}/k$, and $\mathcal{P}_S(k)$ is evaluated upon the horizon exit during inflation. Substituting Eq.~(\ref{eq:psSec}) into Eq.~(\ref{eq:GWenergy}), the fractional energy density of the GWs in the RD era becomes~\cite{Kohri:2018awv, Cai:2018dig, Fu:2019vqc, Lu:2019sti}
\begin{align}\label{eq:GWenergy01}
\Omega_{GW}(k,\tau) = \frac{1}{6} \left( \frac{k}{\mathcal{H}}\right)^2 \int^\infty_0 dv \int^{1+v}_{|1-v|} du \left[ \frac{4v^2-(1-u^2+v^2)}{4uv}\right]^2 \overline{I_{RD}^2(u,v,x)} \mathcal{P}_S(kv)\mathcal{P}_S(ku)\,,
\end{align}
where the full expression of $I_{RD}$ is given by%~\cite{Kohri:2018awv, Cai:2018dig, Fu:2019vqc, Lu:2019sti}
\begin{align}
\overline{I_{RD}^2(u, v, x\rightarrow \infty)} =& \frac{9}{u^2 v^2} \left(\frac{u^2+v^2-3}{2 u v}\right)^4\nonumber\\
&\times \left[\left( \ln\left| \frac{3-(u+v)^2}{3-(u-v)^2}\right|-\frac{4 u v}{u^2+v^2-3}\right)^2+\pi ^2 \Theta \left(u+v-\sqrt{3}\right)\right]\,.
\end{align}
The observed energy densities of GWs today are related to their values after the horizon reentry in the RD era as~\cite{Inomata:2018epa} 
\begin{align}\label{eq:GWES}
\Omega_{GW,0}(k) h^2 =  0.38\,\Omega_{RD, 0}h^2 \left( \frac{g_\ast}{106.75}\right)^{-\frac13}\Omega_{GW}(k, \tau_c)\,,
\end{align}
where $\Omega_{RD,0} h^2\simeq 4.2\times 10^{-5}$ with $h=H_0/100\,\text{km}\,\text{s}^{-1}/\text{Mpc}$ is the fractional energy density of radiation today, and $\Omega_{GW}(k)$ is the GW energy density at the time of horizon reentry of the mode $k$. A relation between the frequency and comoving wavenumber of induced GWs is 
\begin{align}
f = 1.546\times 10^{-15}\left( \frac{k}{\text{Mpc}^{-1}}\right) \text{Hz}\,.
\end{align}
We plot Eq.~(\ref{eq:GWES}) as a function of frequency in the right panel of Fig.~\ref{fig:pbh} together with the sensitivity curves of various GW experiments and missions targeted at different frequency bands. The figure shows that the GWs with the mHz peak frequency are produced if the primordial power spectrum peaks at $k_\text{PBH}=2.45\times 10^{12}\,\text{Mpc}^{-1}$; see the solid black line. The signal of induced GWs falls into the frequency range targeted by the space-borne GW detectors, including LISA~\cite{Audley:2017drz}, Taiji~\cite{Hu:2017mde}, TianQin~\cite{Luo:2015ght}. Similarly, the GWs with the nHz peak frequency can also be produced in the RD era if the primordial power spectrum peaks at $k_\text{PBH}=4.3\times 10^6\,\text{Mpc}^{-1}$; hence, the predictions of our model can be proved by the PTA experiments, including EPTA~\cite{Ferdman:2010xq}, NANOGrav~\cite{McLaughlin:2013ira}, and SKA~\cite{Moore:2014lga}. Our results tell us that different peak scales in the scalar power spectrum correspond to different frequency of GWs; hence, the signals are probed by different experiments. %The $\Omega_\text{GW}(f)h_0^2$ peaks are considerably noticeably broader than that of the scalar power spectra due to the nature of the integrals that determine $\overline{\mathcal{P}_T(k,\tau)}$ in Eq.~(\ref{eq:GWenergy})~\cite{Cai:2018dig}.

%%========================Footnote========================================================================================================================================================
%\begin{figure}[h!]
%\centering  
%\includegraphics[width=0.8\textwidth]{GWspectrum.pdf}  
%\caption{The predicted energy spectrum of the scalar-induced secondary GWs from Eq.~(\ref{eq:GWES}). % where the dashed and solid black lines correspond to modes with $k_\text{PBH}=2\times10^{6}~\text{Mpc}^{-1}$ and $2.45\times10^{12}~\text{Mpc}^{-1}$, respectively. 
%The numerical inputs are the same as Fig.~\ref{fig:pbh} and the colored lines illustrate the sensitivity curves of the current and future GW projects.  
%}\label{fig:gws}
%\end{figure}  
%%=============================---========================================================================================================================================================

\section{Inflaton potential and self-coupling functions}\label{sec:pot}

\begin{figure}[h!]
\centering  
\includegraphics[width=0.49\textwidth]{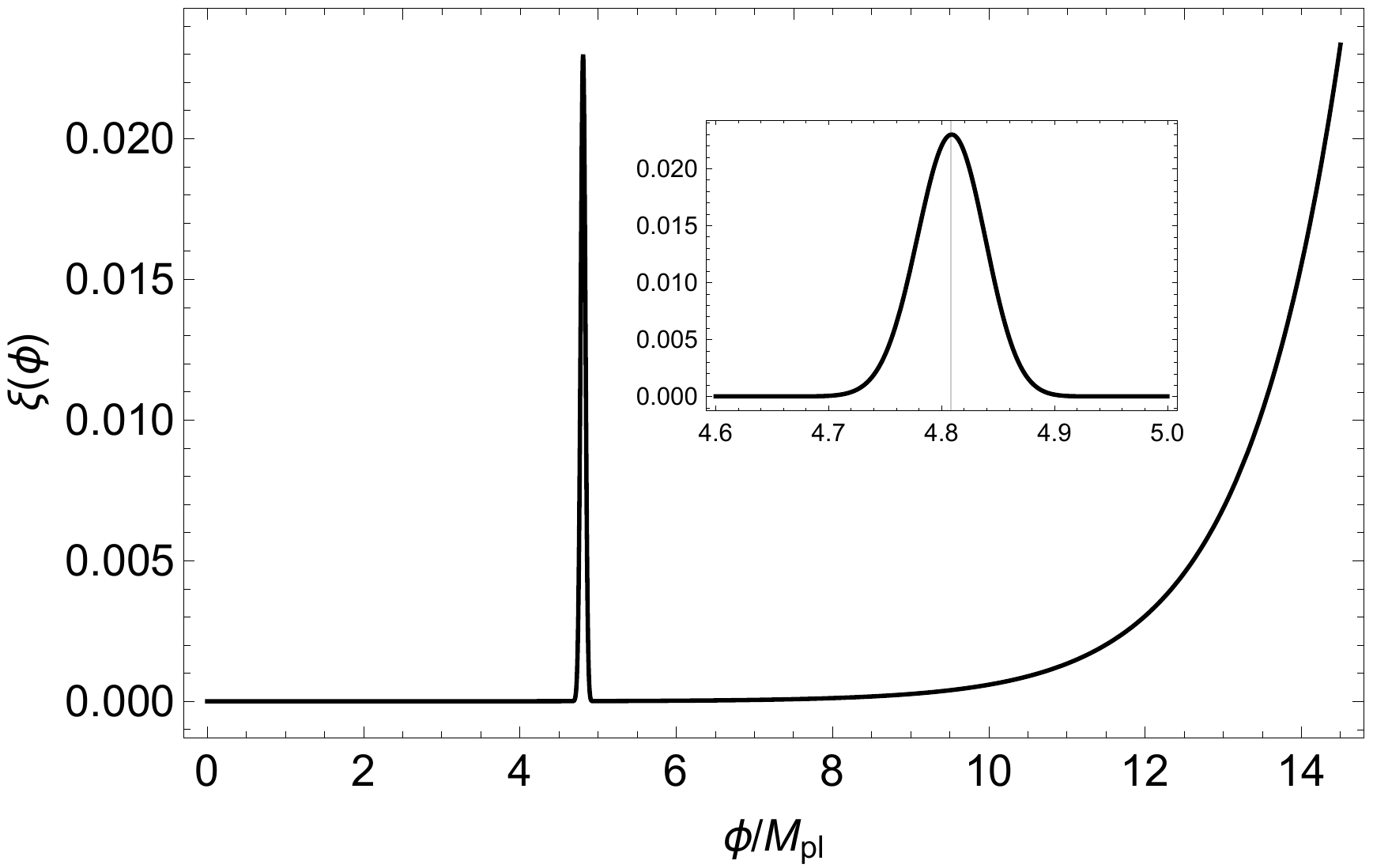} 
\includegraphics[width=0.5\textwidth]{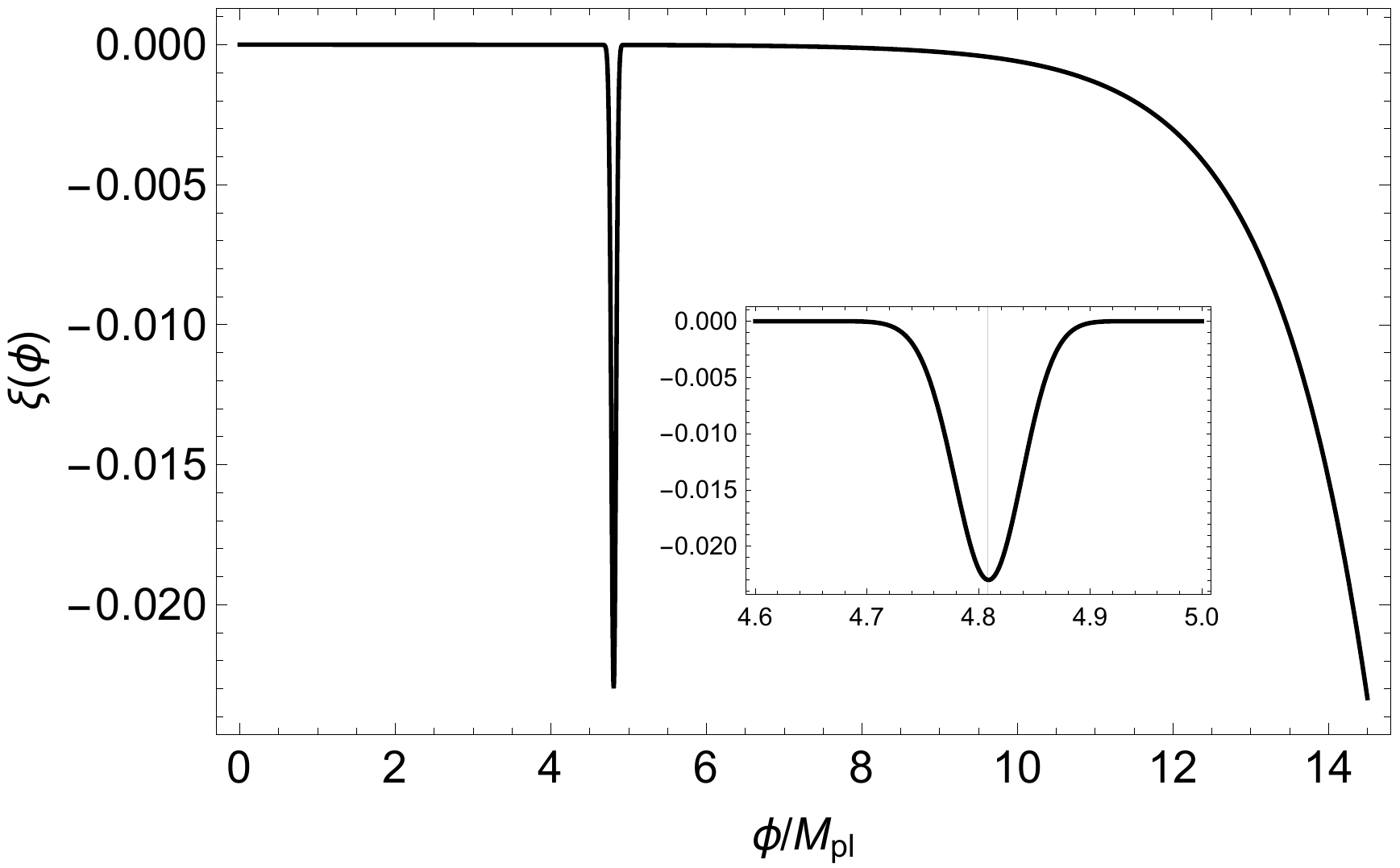} 
\caption{The self-coupling function from Eqs.~(\ref{eq:recxi}) for positive (left) and negative (right) values of $\beta/\alpha$. The peak/dip position $\phi_p \simeq4.8082$ is adjusted by the $k_\text{PBH}$ value. Numerical inputs are $\alpha=\pm 10^3$, $\beta=-10^{-3}$, $\gamma=0.55$, $M=M_{pl}=1$, $\sigma=0.3$, $A_S=6\times 10^{6}$, and $k_\text{PBH}= 4.3\times 10^{6}\text{Mpc}^{-1}$.  } \label{fig:xiphi}
\end{figure}

In Sec.~\ref{sec:setup}, we discussed that the primordial curvature power spectrum could be enhanced due to the self-interaction of the inflaton field and the kinetic coupling between the inflaton and gravity. Then,  by proposing the power spectrum with a local Gaussian peak, we estimated the abundance of PBHs and the energy spectrum of induced GWs in Sec.~\ref{sec:powerspec}. Therefore, it is imperative for us to investigate the configuration of inflaton potential and self-coupling function that yields the power spectrum given in Eq.~(\ref{eq:powerSs}). In order words,  in this section, we reconstruct the inflaton potential and self-coupling function for our model from the power spectrum and its spectral tilt. Our reconstruction method closely follows Refs.~\cite{Chiba:2015zpa}.
We start from Eq.~(\ref{eq:efold}) to obtain 
\begin{align}
\epsilon_V = \frac{1}{2}(1+\mathcal{A})\frac{V_{,N}}{V}\,, \qquad \eta_V= \frac{1}{2}\left(1+\mathcal{A}\right)\left(\frac{V_{,N}^2+VV_{,NN}}{VV_{,N}}+\frac{\left(1+\mathcal{A}\right)_{,N}}{1+\mathcal{A}}\right)\,,
\end{align}
where $V_{,N}>0$ is required. Consequently, we derive from Eqs.~(\ref{eq:nstilt})--(\ref{eq:ttos})
\begin{align}\label{eq:nsofN}
n_S-1 &\simeq  \ln\left[ \frac{V_{,N}}{V^2}\left( 1+\mathcal{A}\right)\right]_{,N}\,, \qquad r=\frac{8V_{,N}}{V}\,. 
%\label{eq:rofN}
\end{align}
Since the potential and couplings are functions of the scalar field, we need a relation between $\phi$ and $N$, which we obtain from Eq.~(\ref{eq:efold}) as 
\begin{align}\label{eq:phiofN}
\int_{\phi_e}^{\phi_\ast} d\phi &=M_{pl}\int \sqrt{\frac{V_{,\tilde{N}}}{V(1+\mathcal{A})}}d\tilde{N}\,,
\end{align}
where $\phi_e$ and $\phi_\ast$ are scalar field values at the end of inflation and the horizon exit of the CMB mode, respectively. Thus, Eqs.~(\ref{eq:nsofN})--(\ref{eq:phiofN}) are the key equations for reconstructing $V(\phi)$ and $\xi(\phi)$.
In this work, we adopt the following form of $n_S(N)$,
\begin{align}\label{eq:nsform}
n_S-1=-\frac{2}{N}\,,
\end{align}
which is in good agreement with the CMB measurement~\cite{Ade:2013zuv} for $N_\ast\simeq 60$ and is extensively studied in the literature~\cite{Chiba:2015zpa}. 
Applying Eq.~(\ref{eq:nsform}) to Eq.~(\ref{eq:nsofN}), we obtain 
\begin{align}
-\frac{2}{N} = \ln\left[ \frac{V_{,N}}{V^2}\left( 1+\mathcal{A}\right)\right]_{,N}%-\frac{\mathcal{A}}{1+\mathcal{A}}\frac{V_{,N}}{V}
\,,
\end{align}
which can also be integrated to give
\begin{align}\label{eq:potN}
\frac{V_{,N}}{V^2}\left( 1+\mathcal{A}\right) = \frac{c_1}{N^2}\,,
\end{align}
where $c_1$ is a positive constant of integration since $V_{,N}>0$.   By comparing Eq.~(\ref{eq:scalarPs}) with Eq.~(\ref{eq:powerSs})  in the $k \gg k_\ast$ limit, we obtain
\begin{align}\label{eq:AasN}
\mathcal{A}(N)%&= \frac{A_S}{\sqrt{2\pi\sigma^2}} e^{-\frac{\ln\left( k/k_\text{PBH} \right)^2}{2\sigma^2} } \nonumber \\
&= \frac{A_S }{ \sqrt{2\pi\sigma^2}} e^{-\frac{\bar{N}^2}{2\sigma^2}}\,,
\end{align}
where $\bar{N}\equiv N-N_p$ is defined from the end of inflation.
%%========================From k to N ========================================================================================================================================================
%~\footnote{\footnotesize{Assuming $H=\text{const.}$ during inflation, one can write in Eq.~(\ref{eq:AasN}): $$\ln\left( \frac{k}{k_\text{PBH}}\right)=\ln\left( \frac{k_\ast}{k_\text{PBH}}\frac{k}{k_\ast}\right)=\ln\left[ \frac{(a H)_\ast}{(a H)_\text{PBH}} \frac{(a H)_k}{(a H)_\ast}\right]=\ln\left(\frac{a_\ast}{a_\text{PBH}}\right)+\ln\left(\frac{a_k}{a_\ast}\right)=N_k-(N_\ast-N_\text{PBH})=N_k-N_p.$$ Here $k_\ast \lesssim k \lesssim k_\text{max}$ and $k_\ast \lesssim k_\text{PBH} \lesssim k_\text{max}$, therefore:
%\begin{itemize}
%\item If $k=k_\ast$, then $N_k=0$ hence $\bar{N}=-(N_\ast-N_\text{PBH}) =-N_p$. 
%\item If $k=k_\text{PBH}$, then $N_k=N_\ast-N_\text{PBH}$ hence $\bar{N}=0$. 
%\item If $k=k_\text{max}$, then $N_k=N_\ast$ hence $\bar{N}=N_\text{PBH}$.
%\end{itemize}%The range of $N_k$ is given by $0\leq N_k\leq N_\ast$.}}
%%========================================================================================================================================================================================= 
The $N$ counts the $e$-folds between the horizon exit of a mode $k$ during inflation and the end of inflation. The difference $N_p=N_\ast-N_\text{PBH}$ indicates how long after the CMB mode $k_\ast$ exit the horizon, the mode $k_\text{PBH}$ corresponding to PBHs would exit the horizon during inflation. Thus, the $N$ varies within an interval $0\leq N \leq N_\ast$ as the $k$ runs between $k_\ast\leq k \leq k_\text{max}$, where the $k_\text{max}$ indicates the smallest scale that exits the horizon during inflation. The CMB data favor $N_\ast\simeq 50 - 60$~\cite{Ade:2013zuv}.  
We solve Eq.~(\ref{eq:potN}) for $V(N)$ using Eq.~(\ref{eq:AasN}), and the solution is given by
\begin{align}\label{eq:potofNnew}
%V(N) = \frac{1+\mathcal{A}(N)}{c_0+c_1/N}\,,
V(N) = \left[c_0+\frac{c_1}{N} - \frac{2c_1\sigma}{\left( 1+ \frac{A_S}{\sqrt{2\pi \sigma^2}}\right)N_p^2}\right]^{-1}\,,
\end{align}
where $c_0$ is an integration constant. The last term in Eq.~(\ref{eq:potofNnew}) is negligible compared to the second term $c_1/N$ if we consider $A_S\sim \mathcal{O}(10^6-10^{7})$, $\sigma\sim\mathcal{O}(0.1-1)$ and $N_p\sim\mathcal{O}(10)$. 
Consequently, from Eq.~(\ref{eq:gammadef}), we obtain
\begin{equation}
\xi(N) = \gamma\frac{\beta M}{\alpha M_{pl}} \sqrt{\frac{V}{V_{,N}}\left(1+\mathcal{A}\right)}\,.
\end{equation}
After using Eq.~(\ref{eq:phiofN}), which allows us to interchange $N$ with $\phi$, we get
\begin{align}\label{eq:recpot}
V(\phi) &= \frac{1}{c_0} \tanh^2\left( \frac{c_2}{2}\frac{\phi}{M_{pl}} \right)\,,\\
\xi(\phi) &= \gamma\frac{\beta M}{2\alpha c_2 M_{pl}} \sinh\left(c_2 \frac{\phi}{M_{pl}}\right)\sqrt{1+ \mathcal{A}(\phi)}
\,,\label{eq:recxi}
\end{align}
where 
\begin{align}
\mathcal{A}(\phi) = \frac{A_S}{\sqrt{2\pi \sigma^2}} e^{-\frac{1}{2\sigma^2}\left[ \frac{1}{c_2^2} \sinh^2 \left( \frac{c_2}{2}\frac{\phi}{M_{pl}}\right) - \phi_p \right]^2}\,,
\end{align}
and $c_2=\sqrt{c_0/c_1}$. The peak in the power spectrum Eq.~(\ref{eq:powerSs}) is, therefore, explained by a local feature of the self-coupling function at $\phi_p = 2/c_2\,\text{arcsinh}\left( \sqrt{c_2^2N_p}\right)$.
The Eqs.~(\ref{eq:recpot}) and (\ref{eq:recxi}) are our key results of this section and describe the shape of the inflaton potential and self-coupling functions, respectively. It is worth noting here that the potential obtained in Eq.~(\ref{eq:recpot}) is the same as that in the so-called ``T-model'' of inflation proposed in Ref.~\cite{Kallosh:2013maa}. In Fig.~\ref{fig:xiphi}, we plot Eq~(\ref{eq:recxi}) as functions of the scalar field $\phi$ for positive and negative values of $\beta/\alpha$.  In conclusion, if the sign of $\beta/\alpha$ is positive (negative), there appears a bump (dip) in the self-coupling function $\xi(\phi)$ at $\phi_p$. The value of $\phi_p$ determines the location of a peak in the power spectra. In other words, the peak scale at which the scalar power spectrum is enhanced can be easily adjusted by the value of $\phi_p$. 

%---------------------------------------------------------------------------------------------------------------------------------------------------------------------------------------------------------------------------------------
\section{Conclusion}\label{sec:conclusion}
%---------------------------------------------------------------------------------------------------------------------------------------------------------------------------------------------------------------------------------------

We have considered the cosmological model Eq.~(\ref{eq:NMDCandGinf}) that incorporates the derivative self-interaction of the scalar field and the kinetic coupling between the scalar field and gravity to investigate the formation of PBHs and induced GWs from inflationary quantum fluctuations. We have derived the inflationary observables and showed in Sec.~\ref{sec:setup} that such additional interactions leave their imprints in the primordial power spectrum hence the spectral tilts, as well as the tensor-to-scalar ratio. In particular, the power spectrum of the scalar perturbations Eq.~(\ref{eq:scalarPs}) is enhanced compared to the case in GR, whereas the tensor spectrum Eq.~(\ref{eq:tensorPt}) is suppressed. 
If the power spectra are enhanced on scales smaller than the CMB pivot scale, they lead to the formation of PBHs and the production of GWs in the RD era. 

The enhancement of the power spectrum is often associated with the local feature of inflaton potentials. However, not every inflaton potentials induce such enhancement; hence a certain degree of fine-tuning is needed. To avoid such fine-tuning, we have specified and proposed the form of the power spectrum in Sec.~\ref{sec:powerspec} instead. The proposed power spectrum is parameterized with a local smooth Gaussian peak, as is seen in Eq.~(\ref{eq:powerSs}) and Fig.~\ref{fig:powerspectrum}. The numerical results of Sec.~\ref{sec:powerspec} are presented in Fig.~\ref{fig:pbh}. Our findings can be summarized as follows. With the enhanced power spectra, PBHs are produced with the peak mass around $10^{-13}-10^{-12}M_\odot$ and $0.1-1M_\odot$. The corresponding peak abundances are $f_\text{PBH}\sim1$ and $f_\text{PBH}\sim 0.05$. As a result, the PBHs with the peak mass around $10^{-13}-10^{-12}M_\odot$ can explain almost all the DM in the universe today, while only up to $5\%$ of the observed DM can be explained by the PBHs with the peak mass around $1M_\odot$. Moreover, the induced GWs with the peak frequency around mHz and nHz can be produced. The induced GWs in our model can be tested by both the space-borne GW detectors and PTA observations. 

To explain the enhancement in the power spectrum, we have reconstructed the inflaton potential and self-interaction functions in Sec.~\ref{sec:pot} from the proposed power spectrum and its spectral tilt. The main results of Sec.~\ref{sec:pot} are, therefore,  Eqs.~\ref{eq:recpot} and (\ref{eq:recxi}). The inflationary predictions of our model are consistent with Planck data~\cite{Ade:2013zuv}. The reconstructed inflaton potential has the same form as the so-called `` T-model'' of inflation~\cite{Kallosh:2013maa}. The enhancement in the power spectrum can be explained by the local feature of the self-coupling function of the inflaton field. Depending on the $\beta/\alpha$ sign, there appears a peak or dip in the small field range of the coupling functions. The precise location of the peak is adjusted by how long after the CMB mode the relevant mode that produces a peak in the power spectra leaves the horizon during inflation.  

In conclusion, for our model, the PBHs and GWs are successfully produced as long as the primordial curvature power spectrum is enhanced on scales smaller than the CMB pivot scale. The enhancement mechanism is then explained by the local feature, either a peak or dip, of the inflaton self-coupling function rather than the local feature of the inflaton potential.

\acknowledgments
PC and GT are supported by Ministry of Science and Technology (MoST) under grant No. 109-2112-M-002-019. SK was supported by the National Research Foundation of Korea (NRF-2016R1D1A1B04932574, NRF-2021R1A2C1005748) and by the 2020 scientific promotion porgram funded by Jeju National University.

%========================================================================================================================================

\end{document}